\documentstyle[12pt]{article}
\newcounter{subequation}[equation]
\makeatletter
\expandafter\let\expandafter\reset@font\csname reset@font\endcsname
\newenvironment{subeqnarray}
  {\arraycolsep1pt
   \def\@eqnnum\stepcounter##1{\stepcounter{subequation}{\reset@font\rm
     (\theequation\alph{subequation})}}\eqnarray}%
  {\endeqnarray\stepcounter{equation}}
\makeatother

\normalbaselines 
\begin{document} 
\begin{center}
\vspace*{1.0cm}

{\LARGE{\bf Generalized Superconformal Symmetries  and 
Supertwistor Dynamics}}\footnote{Invited talk given at the
Conference ``Quantum Theory and Symmetries", Goslar, 18.07-22.07.1999,
presented by J. Lukierski. Supported by KBN grant 2P03.B130.2.}

\vskip 1.5cm

{\large {\bf
Igor Bandos$^{\dag,2}$, Jerzy Lukierski$^{\dag,\ddag}$ and
Dmitri Sorokin$^{\star,}$\footnote{On leave of absence from Institute for 
 Theoretical Physics, NSC Kharkov Institute of Technology,
310108 Kharkov, Ukraine} 
}}

\vskip 0.5cm 

$^{\dag}$Abdus Salam International Centre for Theoretical Physics,\\
Trieste--Miramare, 34100 Trieste, Italy
\\
$^{\ddag}$Institute for Theoretical Physics, University of Wroc\l aw, \\
 pl. Maxa Borna 9, 50-204 Wroc\l aw, Poland
\\
$^{\star}$INFN, Sezione di Padova,
\\
Via F. Marzolo   8, 35131 Padova, Italy

\end{center}

\vspace{1 cm}

  \begin{abstract}
  We show that in the supersymmetry framework described by
a Poincar\'{e} superalgebra with tensorial central charges the
role of generalized superconformal symmetry 
which contains all these central charges is played by $OSp(1|2^{k})$,
where $k=3$ for $D=4$. Following [1,2] we describe the free
supertwistor model for $OSp(1|8)$. It appears  that in such a
scheme the tensorial central charges satisfy additional
relations and the model describes the tower of supersymmetric 
massless states with an arbitrary (integer and half--integer) 
 helicity spectrum.
\end{abstract}

\vspace{1 cm} 

\section{Introduction}
Let us recall that in the standard supersymmetry scheme [3,4] only
scalar central charges are allowed. In particular $D=4$
$N$-extended SUSY has the following most general
form\footnote{For simplicity, we consider here  only the
supercharge sector (``fermion-fermion" relations), which forms
the subsuperalgebra of the full $N$-extended Poincar\'{e} superalgebra.}
 ($A,B=1,2; i,j,k,l=1\ldots N$; $\bar{Q}_{\dot{A}i}=(Q_A^i)^{\dagger }$)
$$
 \left\{ Q_{A}^{i}, \bar{Q}_{\dot{B}j}\right\}
=      \delta^i_{j}(\sigma^{\mu})_{A\dot{B}}P_{\mu}, 
$$
\begin{equation}\label{1}
 \left\{ Q_{A}^{i}, {Q}^{j}_{B}\right\}
  =  \varepsilon_{AB}Z^{[ij]}, \qquad 
 \left\{ \bar{Q}_{\dot{A}i}, \bar{Q}_{\dot{B}j}\right\}
 =   \varepsilon_{\dot{A}\dot{B}}\bar{Z}_{[ij]},
 \end{equation}
 where $Z^{[ij]}= -Z^{[ji]}$ describes   complex scalar 
Abelian central charges
 \begin{equation}\label{2}
 \left[ Q^{i}_{A}, Z^{[kl]}\right] =
 \left[ \bar{Q}^{i}_{\dot{A}}, Z^{[kl]}\right]= 0, \qquad 
\left[ Q^{i}_{A}, P_{\mu}\right] = 0.
 \end{equation}
In particular, if $N=1$ the central  charges are not present, and 
for $N=2$ ($i,j=1,2$) one can introduce one central charge
$Z=\varepsilon_{ij}Z^{[ij]}$.

In a generalized supersymmetry scheme, in order to characterize
all possible sources describing charged $p$-branes, domain walls
etc. one uses nonvanishing tensorial central charges
 $Z^{ij}_{\mu_{1}\ldots \mu_{k}}$, where $k\leq {[D]\over2}$.
For example, in $D=4$ the relations (1) are extended  as follows [5]
\begin{eqnarray}\label{3}
 \left\{ Q_{A}^{i}, \bar{Q}^{\dot{B}j} \right\}
& =& 
      \sigma^{\mu}_{A\dot{B}} P_{\mu} \delta^i_j 
+ i(\sigma^{\mu})_{A\dot{B}} Y_{\mu j}^{i}, 
\cr
 \left\{ Q_{A}^{i}, {Q}^{j}_{B} \right\}
& =& 
\sigma^{\mu\nu}_{A{B}} \widetilde{Z}_{\mu\nu}^{(ij)}
+ \varepsilon_{AB} Z^{[ij]},
\cr
 \left\{ \bar{Q}_{\dot{A}}^{i}, \bar{Q}^{j}_{\dot{B}}\right\}
& =&      \tilde{\sigma}^{\mu\nu}_{\dot{A}\dot{B}} 
\bar{\widetilde{Z}}_{\mu\nu (ij)}
+ \varepsilon_{AB} \bar{Z}_{[ij]},
\end{eqnarray}
where $P_{\mu}$ and  traceless  
 $Y^{i}_{\mu j}$ ($Y^{i}_{\mu i}\equiv 0$) are Hermitian,  
  $ \widetilde{Z}_{\mu\nu}^{(ij)}= - \widetilde{Z}_{\mu\nu}^{(ij)}$
are complex and self--dual with respect to space--time indices, and 
their Hermitian conjugate 
   $\bar{\widetilde{Z}}_{\mu\nu (ij)} =
   \bar{\widetilde{Z}}_{\mu\nu (ji)} = - \bar{\widetilde{Z}}_{\nu\mu (ij)}$
    are anti-selfdual. 
    It should be mentioned that from the algebraic point of view it
is also possible to introduce additional spinorial fermionic 
charges into  $[P,Q]$ and $[Z,Q]$ commutators 
(see e.g. [7,8]). In this talk we shall  consider
only the case of bosonic tensorial central charges.

In particular, for $N=1$, $D=4$,  
using real Majorana supercharges we get 

\begin{equation}\label{5}
\left\{ Q_{\alpha}, Q_{\beta}\right\} = P_{\alpha\beta}
= \left(\gamma^{\mu}C\right)_{\alpha\beta} P_{\mu}
+\left( \sigma^{\mu\nu}C\right)_{\alpha\beta} Z_{\mu\nu}, 
\end{equation}
where the real tensor $Z_{\mu\nu}$ describes six tensorial
central charges. It has been firstly shown in [1]  that the
presence of nonvanishing tensorial central charges $Z_{\mu\nu}$
allows one to describe the superparticle model  with only one
broken target space supersymmetry
 (i.e. 3/4 SUSY remains unbroken)\footnote{See \cite{6} 
for recent discussion of such BPS states.}.
 
If we wish to realize the superalgebra (\ref{5})  as an extension
of the standard superspace framework one should introduce additional 
6 bosonic central charge coordinates (see e.g. [1,2,9]). With increase in
the number of space-time dimensions  the number of these additional
bosonic coordinates grow rapidly (e.g. for $D=11$ one has
 ${32\times 33\over 2} -11 = 517$ central charge coordinates).
However, one can use the twistor and supertwistor framework  \cite{10,11} and
extend the Penrose formula for massless momenta
\begin{equation}\label{6}
P_{A\dot{B}} = \left(\sigma^{\mu}\right)_{A\dot{B}} P_{\mu} 
= \lambda_{A}\lambda_{\dot{B}}
\end{equation}
to the sector of central charges by assuming composite 
formulae\footnote{In \cite{Bars2} tensorial central charges are, in contrast, 
the composites of the momenta for a multiparticle--multitime system.} 
\begin{equation}\label{7}
Z_{AB} = {1\over 4} \sigma^{\mu\nu}_{AB} Z_{\mu\nu} = \lambda_{A}
\lambda_{B}, \qquad
Z_{\dot{A}\dot{B}} = {1\over 4} 
 \sigma^{\mu\nu}_{\dot{A}\dot{B}} Z_{\mu\nu} = \lambda_{\dot{A}}
 \lambda_{\dot{B}}. 
\end{equation}
In such a way we obtain the tensorial central
charges satisfying some constraints, and the number of
independent degrees of freedom in dimension $D$      is
determined by the real dimension of the fundamental spinor
representation. In particular, for $D=4$ we have $n=4$
 degrees of freedom, for $D=5,6$ and 7\, there are  $n=8$  degrees, for
$D=8,9$ and 10\,  $n=16$, and for $D=11$ we have $n=32$. 
If we consider the dimensions $D=3,4,6,10$
describing respectively the sequence of real, complex,
quaternionic and octonionic 
  pair of supercharges we see that subtracting $D-1$
degrees of freedom describing massless momenta we obtain
respectively $m=0,1,3,7$ internal variables parametrizing
the spheres $S^{m}$ \cite{2}. 

\section{Generalized Superconformal Symmetries 
 and Tensorial Central Charges}

Standard conformal algebras in $D$ dimensions are isomorphic to 
the orthogonal algebras $O(D,2)$. In order to introduce the
standard conformal superalgebra we should consider the fundamental
spinorial realization of $O(D,2)$. The
dimensions $D=3,4$ and 6 are specific in the sense that then the 
spin coverings of $SO(D,2)$ 
are described by the classical groups, namely,
${\rm Spin}(3,2)={\rm Sp}(4;R)$,
 ${\rm Spin}(4,2)=SU(2,2)$ and ${\rm Spin}(6,2)=U_{\alpha}(4,H)=
 U(4,4)\cap {\rm Sp}(8;C) =USp(4,4;C)$\footnote{We shall not
include into this sequence $D=10$ with octonionic
nonassociative group-like matrices $U_{\alpha}(4;O)$ (for more
extensive discussion see \cite{12}). 
By $U_{\alpha}(N;F)$ we denote the group of symplectic--unitary matrices 
($U^{\dagger} \Omega U= \Omega, ~~\Omega^T = -\Omega$) over a field $F$.}. 
Supersymmetrization implies that
\begin{equation}\label{8}
\begin{array}{llllll}
\displaystyle D &= 3 &: \quad &Sp(4;R) &\to & OSp(N;1|R)
\cr
\displaystyle 
D &= 4 &: \quad &SU(2,2) &\to & SU(2,2|N)
\cr
\displaystyle 
D &= 6 &: \quad &U_{\alpha}(4;H) &\to & UU_{\alpha}(N;4|R)
\end{array}
\end{equation}
The standard conformal algebra $O(D,2)$ has the following three-fold grading
\begin{equation}\label{9}
O(D,2) = P\oplus L \oplus K
\end{equation}
where $P$ corresponds to the sector of $D$ translation generators,
$L=M\oplus R$ contains Lorentz  $O(D-1,1)$ rotations $M$ and the
dilatation generator $R$, and $K$ describe $D$ conformal
boosts. The supersymmetrization of the graded
structure (\ref{9}) for $N=1$ is performed by 
taking the ``supersymmetric square roots" of
the $O(D,2)$ generators in the following way ($G=U(1)$ for
$D=4$, and $U(2)$ for $D=6$):
\begin{eqnarray}\label{10}
\left\{ Q, Q \right\} \subset P , \qquad 
\left\{ S, S \right\} \subset K ,
\qquad
\left\{ Q, S \right\} \subset L \oplus G .
\end{eqnarray}
where $G$ describe internal symmetry generators. 
The superalgebra (\ref{9},\ref{10}) is endowed with a 
five--fold graded structure
$P\oplus Q\oplus(L\oplus G)\oplus S  \oplus K$.

In order to introduce generalized conformal superalgebras
which include tensorial central charges we should generalize the
set of relations (\ref{10}). In $D=4$ the generalized conformal algebra
is obtained by replacing four-momenta generators $P_{\mu}$
with ten generators 
$(P_{\mu},Z_{\mu\nu})$, (see (\ref{5})), and for $D=4$ in place of
(\ref{10}) we obtain the following simple real superalgebra:
 \begin{equation}\label{11}
\left\{ Q_{\alpha}, Q_{\beta} \right\} = {\cal P}_{\alpha\beta}, 
\quad
\left\{ S_{\alpha}, S_{\beta} \right\} = {\cal K}_{\alpha\beta},
\quad
\left\{ Q_{\alpha}, S_{\beta} \right\} = {\cal L}_{\alpha\beta}.
\end{equation}
The relations (\ref{11}) contain two copies of the
Poincar\'{e} superalgebra (\ref{5}) and 16 real bosonic
charges ${\cal L}_{\alpha\beta}$ generate 
$GL(4;R)$. We thus get a
{\it\bf  generalized $D=4$ conformal algebra given by $Sp(8)$}
(note that $Sp(8)$ contains $SU(2,2)$)
\begin{equation}\label{12} 
SU(2,2) = Spin(4,2) \qquad \subset \qquad 
Sp(8)= {\cal P} \oplus {\cal L} \oplus {\cal K}, \quad 
\end{equation}
and the generalized $D=4$ superconformal algebra (\ref{11}) is
described by $OSp(1;8)$. 

One can also introduce  an N-extended $D=4$ generalized superconformal group  
$ OSp(N;8)$ ($SU(2,2|N) \subset OSp(2N;8)$ as well as the ones for $D > 4$ 
containing D--dimensional super--Poincar\'{e} 
algebra with all possible tensorial charges \cite{13,Bars1}. 
 For
example, in $D=10$ all ${126}$ central charges are included in the simple 
generalized superconformal algebra $ OSp(1;32)$ \cite{Bars1}.

\section{Supertwistorial Realizations 
 of Generalized Superconformal Symmetries}

Let us consider the following realization of the superalgebra
$OSp(1;8)$ described  by the relations (\ref{11}):

  a) Bosonic sector

\setcounter{equation}{11}
\begin{subeqnarray}\label{13a}
P_{\alpha\beta} = \lambda_{\alpha} \lambda_{\beta}, \qquad 
M_{\alpha\beta} = \lambda_{\alpha} \mu_{\beta}, \qquad 
K_{\alpha\beta} = \mu_{\alpha} \mu_{\beta}.
\end{subeqnarray}

b) Fermionic sector ($\xi^{2}=1)$
\setcounter{equation}{11}
 \begin{subeqnarray}\label{13b}
  \setcounter{subequation}{1}
Q_{\alpha} = \lambda_{\alpha} \xi , \qquad S_{\alpha} =
\mu_{\alpha} \xi
\end{subeqnarray}
where 
\setcounter{equation}{11}
 \begin{subeqnarray}\label{13c}
   \setcounter{subequation}{2}
\left\{ \lambda_{\alpha}, \mu^{\beta}\right\} =
\delta_{\alpha}^{\ \beta},
\qquad \xi^{2}=1
\end{subeqnarray}
and raising of indices is performed with the help of an $Sp(4)$
antisymmetric metric.

It is easy to see that the first relation 
(12a) 
is equivalent to the relations (\ref{6}-\ref{7}).

The relation with the superspace description is obtained via the
following version of Penrose-Ferber formulae (see \cite{11}) relating
supertwistor coordinates with superspace coordinates
\begin{equation}\label{14}
   \mu^{\beta} = \lambda_{\alpha} \left( X^{\alpha \beta}
   - i\theta^{\alpha}\theta^{\beta}\right), 
\qquad 
   \xi =\lambda_{\alpha}\theta^{\alpha}
   \end{equation}
   where $(\lambda_{\alpha}, \mu^{\alpha},\xi)$ describes
an $OSp(1|8)$ supertwistor and 
\begin{equation}\label{15}
X^{ \alpha\beta}= {1 \over 4} \left(\gamma_{\mu}C\right)^{\alpha\beta}
x^{\mu} - {1 \over 8}  \left(\sigma_{\mu\nu}C\right)^{\alpha\beta} 
y^{[\mu\nu]}.
\end{equation}
The four coordinates $x^{\mu}$ describe $D=4$ space-time, and
$y^{[\mu\nu]} = -y^{[\nu\mu]}$ are six central charge coordinates,
dual to the tensorial central charges $Z_{[\mu\nu]}$.

\section{New Class of  Massless Superparticle Models}
Our aim here is

i) to construct the superparticle model with the momenta and
tensorial charges given by the relations (\ref{6}, \ref{7}) and 

ii) to show that it is equivalent to the free particle model in
supertwistor space.

We start with the following Brink--Schwarz--like action (see also \cite{14}):
\begin{equation}\label{16}
S = \int d\tau \left( P^{\alpha\beta} \omega_{\alpha\beta} - e
P_{\alpha\beta} P^{\alpha\beta}\right),
\end{equation}
where $P^{\alpha\beta}$ is a symmetric $4\times 4$ matrix, $e$
is an einbein and 
\setcounter{equation}{14}
 \begin{subeqnarray}\label{16a}
   \setcounter{subequation}{0}
   \omega_{\alpha\beta} =  {dX^{\alpha\beta}\over d\tau}
   - i \theta^{(\alpha} {d\theta^{\beta)}\over d\tau}.
   \end{subeqnarray}
Substituting the ansatz (\ref{13a}a) $P_{\alpha \beta} = 
\lambda_{\alpha} \lambda_{\beta}$ 
into (\ref{16}) we get
\begin{eqnarray}\label{17}
S = \int d\tau \lambda^{\alpha}\lambda^{\beta} \omega_{\alpha\beta}
= \int d\tau\left( \lambda^{A}\bar{\lambda}^{B} \omega_{A\dot{B}}
+
\lambda^{A}{\lambda}^{\dot{B}} \omega_{A{B}}
 + \lambda^{\dot{A}}{\lambda}^{\dot{B}} \omega_{\dot{A}\dot{B}}\right)
\end{eqnarray}
where (we use the 2-component Weyl spinor notation) 
\begin{eqnarray}\label{18}
\omega_{A\dot{B}}& = & {dx_{A\dot{B}}\over d\tau }+
i\left( {d\theta_{A}\over d\tau } \bar{\theta}_{\dot{B}} 
- \theta_{A} { d\theta_{\dot{B}}\over d\tau} \right), 
\cr
 \omega_{A{B}}& = & {dy_{A{B}}\over d\tau} -
i {d\theta^{(A}\over d\tau } {\theta}^{{B)}}, 
\qquad \qquad
  \omega_{\dot{A}\dot{B}} = {dy_{\dot{A}\dot{B}}\over d\tau} -
i {d\bar{\theta}^{(\dot{A}}\over d\tau } \bar{\theta}^{\dot{B)}}. 
\end{eqnarray}

The model (\ref{17}) describes the generalized momenta
$P_{\alpha\beta}$ satisfying the relations
$P_{\alpha\beta}P^{\beta\gamma} = 0$ or more explicitly (using
the notation of eqs. 
(\ref{6}-\ref{7})):
\begin{equation}\label{19}
P_{A\dot{B}} P^{A\dot{B}} = 0, \quad
Z_{AB}Z^{AB} = \bar{Z}_{\dot{A}\dot{B}} 
\bar{Z}^{\dot{A}\dot{B}} = 0,
\quad 
Z_{AB}P^{B\dot{C}} = P_{A\dot{B}}Z^{\dot{B}C} = 0
\end{equation}
The relations (\ref{19}) reduce 10 real degrees of freedom ($P_{\mu},
Z_{[\mu\nu]}$) to four real independent  degrees of freedom. In
particular, if we introduce three degrees  describing
$D=4$ massless momenta ($\vec{p}, p_{0}=|\vec{p}|$), one can
describe the 
fourth degree of freedom $e^{i\alpha}$ as the phase of the spinor
$\lambda_{A}$, which e.g. can be expressed as 
$e^{i4\alpha} = {Z_{12}/ \bar{Z}_{\dot{1}\dot{2}}}$

In such a way we obtain the $D=4$ massless superparticle model
with additional  ``internal" $U(1)$ degree of freedom. In
particular if we perform quantization (see \cite{2}) we obtain the
superwave function $ \Phi$ which depends only on independent
variables $\lambda_{A},\bar{\lambda}_{\dot{A}}$ and
one-dimensional Grassmann coordinate $\eta$ ($\eta^{2}=0$; see  (\ref{14}))
\begin{equation}\label{21}
\Phi\left( \lambda_{A}, \bar{\lambda}_{\dot{A}},\eta\right)
= \Phi
\left( \lambda_{A}, \bar{\lambda}_{\dot{A}}\right)
+ i\eta \Psi \left(
\lambda_{A}, \bar{\lambda}_{\dot{A}}\right)
\end{equation}
Since 
the set of variables $\left(\lambda_{A}, \bar{\lambda}_{\dot{A}}\right)$ 
is equivalent to $\left( p_{\mu} ~~(p^{2}=0); ~ e^{i\alpha} \right)$, where
$p_{\mu} = 
{\lambda}
\sigma_{\mu}\bar{\lambda}$, 
and  $\lambda_{1}\lambda_{2}= |\lambda_{1}| |\lambda_{2}| e^{2i\alpha}$, 
  one gets
\setcounter{equation}{19}
 \begin{subeqnarray}\label{23}
   \setcounter{subequation}{0}
\Phi\left( \lambda_{A},\bar{\lambda}_{\dot{A}}\right) =
\sum\limits_{k\in Z}
e^{2ki\alpha} \Phi_{k} (p_{k})
\\ \cr
\Psi\left( \lambda_{A},\bar{\lambda}_{\dot{A}}\right) =
\sum\limits_{k\in Z}
e^{(2k+1)i\alpha} \Psi_{k+{1\over 2}} (p_{k})
\end{subeqnarray}
The massless fields collected in (\ref{23}a) carry 
integer helicities  $(s=k)$, and the fields in (\ref{23}b) 
are endowed with half-integer
helicities $(s=k+{1\over 2})$ in accordance  with the spin-statistics 
theorem for the $D=4$ relativistic theories.

In $D =4$ the model (\ref{17})  contains superspace variables ($X^{\mu},
\theta^{A},\bar{\theta}^{\dot{B}}$) extended by  central charge coordinates 
$y^{\mu\nu}$ 
 as well as by the spinors
$\lambda_{A}, \lambda_{\dot{B}}$ describing half of the bosonic
components of the $OSp(1|8)$ supertwistor. If we substitute the
relations (\ref{14})  adapted to $D =4$ 
into the action (\ref{17}) the latter can be expressed in terms of
supertwistor components $Z_{k}= (\lambda_{A}, \mu^{\dot{A}}, \xi)$ 
and thus  becomes the free $OSp(1;8)$ supertwistor action 
(see \cite{2} for details)
\begin{equation}
S = \int d\tau \left( \lambda_{A}\dot{\bar{\mu}}^{A} +
\bar{\lambda}_{\dot{A}}\dot{{\mu}}^{\dot{A}} + i \xi \dot{\xi}\right)
\end{equation}
where $\xi ={1\over 2} \left( 
\lambda_{A}\theta^{A}  + 
\bar{\lambda}_{\dot{A}} \bar{\theta}^{\dot{A}} \right)$.

\section{Final Remarks}
The classical and quantum version of $D=4$ massless
superparticle model with infinite spectra of helicities can be

 - extended to dimensions $D>4$, in particular to $D=6$ and $D=10$ [2]
 
 - generalized to the super-Anti-de-Sitter background 
\cite{15}.
 \section*{Acknowledgments}
 One of the authors (J.L.) would like to thank prof. H.D.
Doebner and V.D. Dobrev for warm hospitality at the conference
in Goslar.

\end{document}